\documentclass[reprint,
superscriptaddress,
 amsmath,amssymb,
 aps,
 prapplied
]{revtex4-2}

\usepackage{graphicx}% Include figure files
\usepackage{dcolumn}% Align table columns on decimal point
\usepackage{bm}% bold math
\usepackage[dvipsnames]{xcolor}
\begin{document}

\preprint{APS/123-QED}

\title{Electronic \textit{g}-factor and tunable spin-orbit coupling in a gate-defined InSbAs quantum dot}

\author{S. Metti}
\affiliation{Elmore Family School of Electrical and Computer Engineering, Purdue University, West Lafayette, Indiana 47907, USA}
\affiliation{Birck Nanotechnology Center, Purdue University, West Lafayette, Indiana 47907, USA}
\author{C. Thomas}
\affiliation{Department of Physics and Astronomy, Purdue University, West Lafayette, Indiana 47907, USA} 
\affiliation{Birck Nanotechnology Center, Purdue University, West Lafayette, Indiana 47907, USA}
\author{M. J. Manfra$^{\dagger}$}
\affiliation{Department of Physics and Astronomy, Purdue University, West Lafayette, Indiana 47907, USA} 
\affiliation{Birck Nanotechnology Center, Purdue University,  West Lafayette, Indiana 47907, USA}
\affiliation{Elmore Family School of Electrical and Computer Engineering, Purdue University, West Lafayette, Indiana 47907, USA}
\affiliation{School of Materials Engineering, Purdue University, West Lafayette, Indiana 47907, USA}
\affiliation{Microsoft Quantum Lab West Lafayette, West Lafayette, Indiana 47907, USA}

\date{\today}% It is always \today, today,

\begin{abstract}
We investigate transport properties of stable gate-defined quantum dots formed in an InSb$_{0.87}$As$_{0.13}$ quantum well. High \textit{g}-factor and strong spin-orbit-coupling make InSb$_x$As$_{1-x}$ a promising platform for exploration of topological superconductivity and spin-based devices. We extract a nearly isotropic in-plane effective $\textit{g}$-factor by studying the evolution of Coulomb blockade peaks and differential conductance as a function of the magnitude and direction of magnetic field. The in-plane $\textit{g}$-factors, $|g^*_{[1\bar{1}0]}|$ and $|g^*_{[110]}|$, range from 49 - 58. Interestingly, this $\textit{g}$-factor is higher than that found in quantum dots fabricated from pure InSb quantum wells \cite{kulesh2020quantum}. We demonstrate tunable spin-orbit-coupling by tracking a spin-orbit-coupling mediated avoided level crossing between the ground state and an excited state in magnetic field. By increasing the electron density, we observed an increase in an avoided crossing separation, $\Delta_{SO}$. The maximum energy separation extracted is $\Delta_{SO}$$\sim$100 $\mu$eV.   
\end{abstract}

\maketitle

\section{\label{sec:level1} Introduction}
The two-dimensional electron gas (2DEG) confined in InSbAs quantum wells is a promising candidate for exploration of topological superconductivity \cite{Lutchyn.2010, Oreg.2010} and spin-based device applications \cite{datta1990electronic,loss1998quantum} due to its strong spin-orbit-coupling (SOC) \cite{Moehle2021, Metti2022,Sestoft2018} and high effective \textit{g}-factor \cite{mayer2020superconducting,Moehle2021,winkler2003spin}. InSbAs is readily proximitized by s-wave superconductors \cite{mayer2020superconducting, Moehle2021}, rendering it an interesting alternative to binary InSb in structures requiring the marriage of superconductivity and SOC. Furthermore, the 2DEG platform offers advantages for scalability of complex devices. A few experimental studies have begun to explore the benefits of this material system in devices, but many fundamental properties in confined geometries are not yet explored. Rashba SOC strength in planar InSbAs 2DEGs as a function of arsenic mole fraction was explored in surface layers and buried quantum wells \cite{Moehle2021, Metti2022}. Measurement of the \textit{g}-factor has been performed on planar InSbAs/InSb superlattices \cite{jiang2022giant}, high-quality InSbAs alloys \cite{jiang2023g},  and in surface layers \cite{Moehle2021}. While recent studies have shown the promise of this material for the study of topological superconductivity and Majorana chains \cite{mayer2020superconducting, Moehle2021, wang2023triplet, prosko2023flux}, there has been limited work detailing the modification of basic electronic properties in confined low-dimensional structures.

The present work explores the properties of gate-defined quantum dots (QDs) in an InSbAs quantum well. QDs may be used as fundamental building blocks for a variety of quantum device applications \cite{wang2023triplet,prosko2023flux}, enabling quantum simulation \cite{pouse2023quantum}, and spin qubits \cite{loss1998quantum}. Specifically, QDs formed in materials with strong SOC may be employed for readout of topological qubits \cite{plugge2017majorana, karzig2017scalable}, and for spin-orbit controlled spin qubits \cite{nadj2010spin,nowack2007coherent} as the strong SOC enables fast spin manipulation. By employing a local oscillating electric field, an induced spin resonance can be achieved by coupling the field and the spin via the SOC. When tuning a single dot in an array, electrostatic gating has benefits for local manipulation over a static magnetic field and an electron-spin resonance pulse \cite{nadj2010spin,golovach2006electric}. 

Here, we present a detailed study of the operation and properties of a quantum dot defined in an InSb$_{0.87}$As$_{0.13}$ quantum well. We extract the in-plane \textit{g}-factor and examine the impact of SOC on electronic structure in the quantum dot. We extract the in-plane \textit{g}-factor by studying the evolution of Coulomb blockade peaks and differential conductance with magnetic field $B_x \parallel$ [1$\bar{1}$0] and $B_y$ $\parallel$ [110] crystallographic directions. The measurements yield an in-plane isotropic \textit{g}-factor with a value ranging from $|g^*_{\parallel}|\sim$ 49 - 58, significantly higher than $|g^*_{\parallel}|\sim$ 26-35 found in QDs formed in pure InSb quantum wells \cite{kulesh2020quantum}. Additionally, by employing a dual-layer gate design, we present evidence of a tunable SOC through the observation of an avoided crossing between the ground state and an excited state in a magnetic field. This enables the determination of spin-orbit interaction strength by extracting an energy gap for the anti-crossing of $\Delta_{SO} $ of $\sim $100 $\mu$eV.

\begin{figure*}
\includegraphics[width=1.0\textwidth]{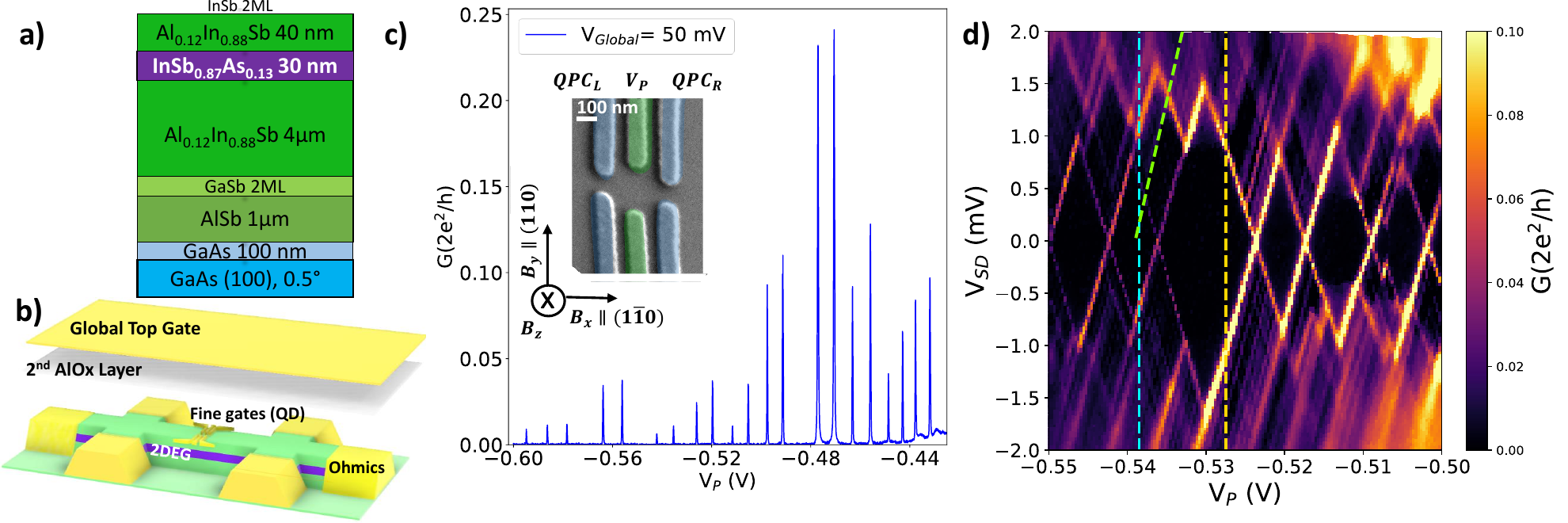}
\caption{
    a) Schematic of the heterostructure employed in this study. 
    b) Model of the device design with dual layer gates insulated with 20 nm Al$_2$O$_3$ (not shown for clarity) below the fine gates  and 40 nm Al$_2$O$_3$ below the global top gate. The first layer of fine gates defines the quantum dot, while the second layer serves as a global gate to control the carrier density. 
    c) Coulomb blockade peaks as a function of plunger gate voltage with AC excitation voltage of $V_{AC}$ = 25 $\mu$V, the tunneling barriers are $QPC_L$ = -0.155 V and $QPC_R$ = -0.185 V, and the global top gate is set to $V_{Global}$ =  +50 mV. Inset: False color SEM image of the quantum dot design and magnetic field directions. The blue gates are the tunneling barriers $QPC_L$ and $QPC_R$,  and the green gates are the plunger $V_P$.
    d) Conductance through the quantum dot as a function of plunger voltage ($V_P$) and source-drain bias ($V_{SD}$).
}  
\label{Fig:fig1}
\end{figure*}

\section{\label{met} Experimental Details}
The heterostructure employed in this work is grown by molecular beam epitaxy on a GaAs (001) substrate with 0.5$^{\circ}$
miscut towards (111)B. The active region is formed by a 30 nm InSb$_{x}$As$_{1-x}$ quantum well with an antimony mole fraction of x = 0.87. The 2DEG is buried 40 nm below the surface, as seen in Fig. \ref{Fig:fig1}(a), and has a zero-bias ($V_{Global}$ = 0 V) 2DEG density of $n$ = 1.5 $\times$ 10$^{11}$ cm$^{-2}$. More details on the heterostructure growth and 2DEG properties may be found in Ref. \cite{Metti2022}. 
\begin{figure*}
\includegraphics[ width=\textwidth]{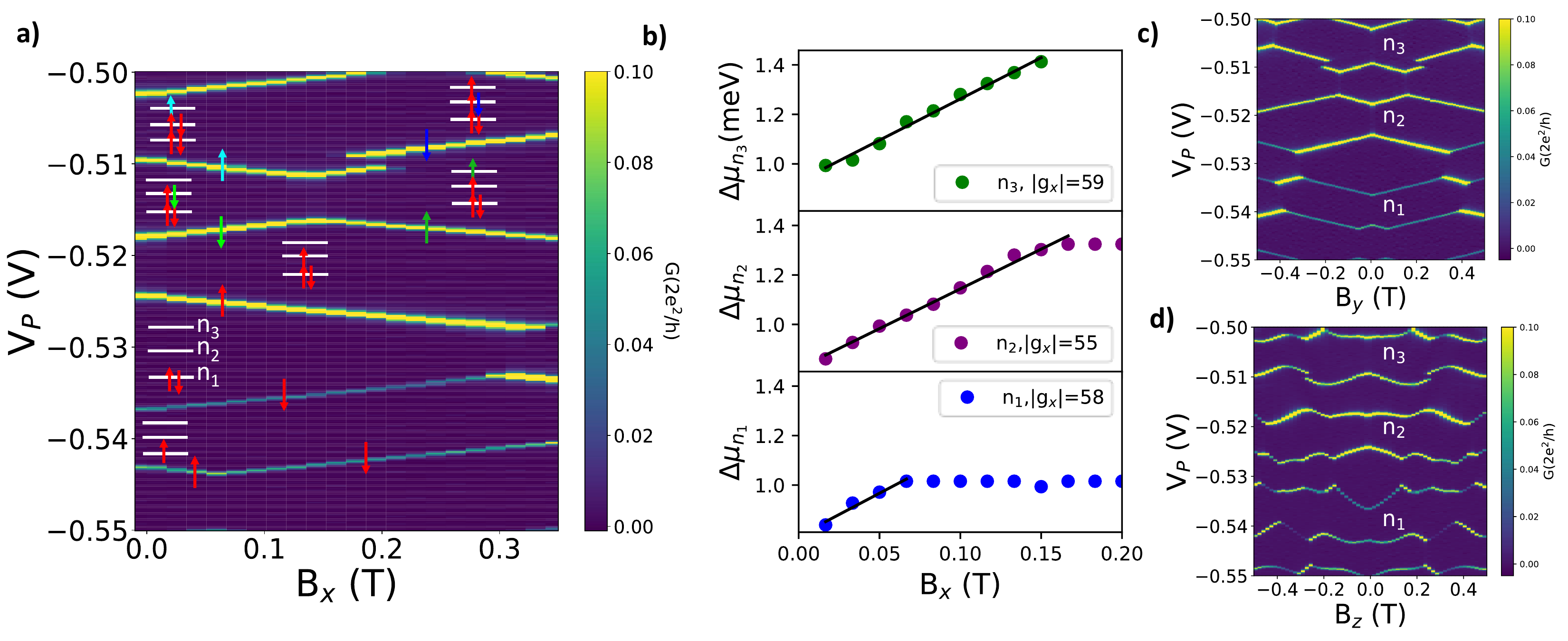}
\caption{
    a) Evolution of the Coulomb blockade peaks as a function of the in-plane magnetic field $B_x$ aligned along the [1$\bar{1}$0] direction. The arrows along the conductance peaks represent the spin of the electron tunneling into a given orbital energy level. The white lines are dot energy levels with the spin configuration specified. The AC excitation voltage is $V_{AC}$ = 35 $\mu$V, while the tunneling barriers and global gate voltage were set to $QPC_L$ = -0.155 V and $QPC_R$ = -0.181 V and  $V_{Global}$ =  +50 mV. 
    b) Addition energy as a function of in-plane magnetic field. The \textit{g}-factor is estimated with a linear fit at low field (black line).
    c) Evolution of the Coulomb blockade peaks as a function of the in-plane magnetic field $B_y$  aligned along the [110].
    d) Evolution of the Coulomb blockade peaks as a function of perpendicular magnetic field $B_z$ aligned along the [001] direction.}
    
\label{Fig:fig2}
\end{figure*}
A QD is fabricated using a dual-layer gate process on top of a mesa forming a Hall bar with ohmic leads. The mesa is 30 $\mu$m wide and 405 $\mu$m long, while ohmic probes are separated by 315 $\mu$m. The dual-layer gate design allows modulation of the electron density by the global top gate once the confined quantum dot geometry is defined by the fine gates.

Preliminary magnetotransport characterization was performed at $T$ = 1.8 K to extract 2DEG carrier density as a function of $V_{Global}$ and test functionality of fine gates. The QD was subsequently measured in a dilution refrigerator at $T$ = 10 mK using low-frequency AC lock-in techniques with AC excitation ($V_{AC}$) of  35 $\mu$V or lower. The QD dual-layer gate design is shown in Fig. \ref{Fig:fig1}(b). The first layer of fine gates establishes the QD, as shown in the inset of Fig. \ref{Fig:fig1}(c); it is composed of two quantum point contacts (QPC) and a pair of middle plunger gates. The left ($QPC_L$) and right ($QPC_R$) gates functioned as the tunneling barriers, and the middle gates functioned as the plunger gate ($V_P$), tuning the dot's size. A second large-area metal layer formed the global gate ($V_{Global}$), which allowed for electrostatic control of the number of electrons in the quantum dot. 

All fine gates ($QPC_L$, $QPC_R$, and $V_P$) were biased cooled from room temperature with a voltage of +0.65 V as bias cooling resulted in more stable device performance. As a first step, the dot was tuned to the Coulomb blockade regime, with $V_{Global}$ set to +50 mV, $V_{AC}$ = 25 $\mu$V, while $QPC_L$ = -0.155 V and $QPC_R$ = -0.181 V . As shown in Fig. \ref{Fig:fig1}(c), we observed transmission peaks indicating the tunneling of electrons in and out of the dot; each peak occurs when the electrochemical potential of the source or drain aligns with one of the dot's unoccupied energy levels \cite{PhysRevB.44.1646}.

With $V_{Global}$ = +50mV resulting in a 2DEG density $n = 1.6\times10^{11}$cm$^{-2}$ and using the lithographic area of the dot, we set an upper bound on the number of electrons (N) to be $\leq$ 90. However, due to lateral depletion around the gates, the electrostatic dimensions of the dot are actually smaller; we roughly estimate a few tens of electrons occupy the dot. For subsequent analysis the magnetic field was aligned parallel to either the [1$\bar{1}$0] or [110] crystallographic directions, as seen in the inset in Fig. \ref{Fig:fig1}(c)

We measured the differential conductance as a function of plunger gate voltage while applying the source-drain bias symmetrically across the device.  $V_{SD}$/2 is applied to the source while -$V_{SD}$/2 is applied to the drain. We set an AC excitation voltage to $V_{AC}$= 35$\mu$V, and the tunneling barriers to $QPC_L$ = -0.155 V and $QPC_R$ = -0.181 V. Well-defined Coulomb diamonds are observed as seen in Fig. \ref{Fig:fig1}(d). The measurements yielded symmetric Coulomb diamonds, which implies equal coupling of the source and drain to the dot. We also observed clear signs of transport through excited states by resolving conductance peaks parallel to the diamond edges. 

 Using standard analysis techniques for QDs \cite{ihn2009}, the plunger gate lever arm was calculated from the Coulomb blockade diamonds shown in Fig. \ref{Fig:fig1}(d). We note that $\alpha_P=\delta V_{SD}$/$ \delta$ $V_P$ = 132 meV/V, where $\delta V_{SD}$ is the addition energy inferred from the diamond's height, corresponding to the electrochemical potential difference between the source and the drain, while $\delta $$V_P$ is the change in plunger gate voltage between the two consecutive Coulomb peaks of the diamond at zero source-drain bias. This calculation was performed for every diamond shown in Fig.\ref{Fig:fig1}(d) to extract an average lever arm. With knowledge of $\alpha_P$, the lever arms of the other gates were extracted by measuring gate-gate maps of pairs of gates. The extracted level arms for each gate are: $\alpha_{Global}$ = 223 meV/V, $\alpha_L$ = 88 meV/V, $\alpha_R$ = 124 meV/V. $\alpha_R$$\ge \alpha_L$ indicates the electron distribution may be slightly shifted towards $QPC_R$. The gate-gate maps of conductance resonances yield lever arms that indicate good electrostatic control of the dot and the absence of accidental dot formation resulting from disorder.  
\break
\indent From the smallest diamond in Fig. \ref{Fig:fig1}(d), we estimated a charging energy of $E_c$ $\geq$ 0.750 meV. We did not observe a systematic variation of the addition energy as a function of added electrons, as may be expected for an even-odd variation resulting from the spin degeneracy of the orbital levels \cite{kulesh2020quantum} or an atom-like orbital filling pattern \cite{kouwenhoven2001few}. The lack of a regular pattern in the diamond size may be associated with strong SOC modification of level spectrum \cite{governale2002quantum} or asymmetry in the confinement potential of the dot \cite{Akbar2001, bjork2004few}. In our case, both mechanisms are potentially active due to InSbAs having a large intrinsic Rashba coupling and the lithographic elliptical shape of the dot. 

 \begin{figure}[b]
    \includegraphics[width=\columnwidth, scale=1.5
]{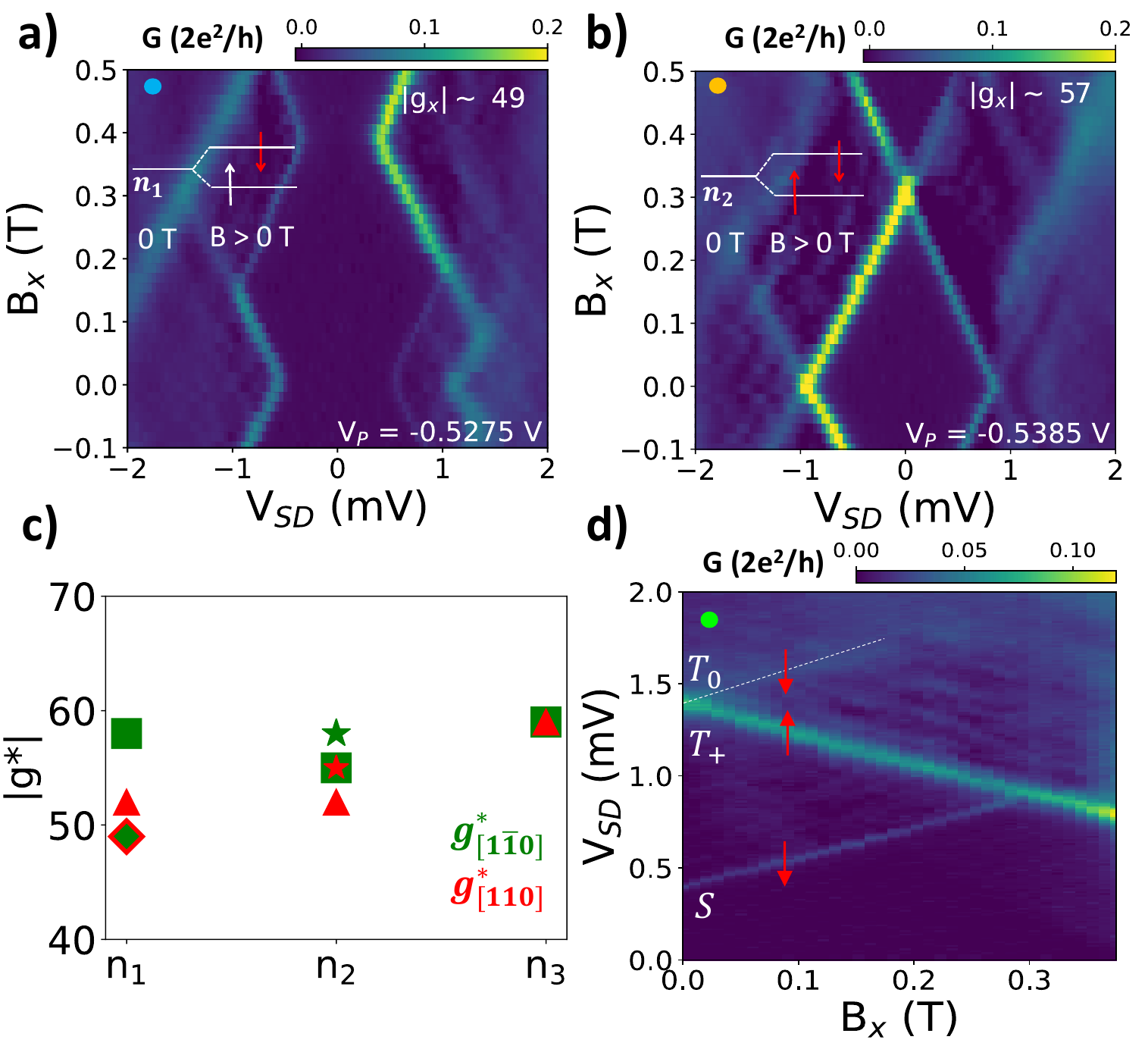}
\caption{
    a) Evolution of the differential conductance with fixed plunger gate voltage, $V_P$ = -0.5275 V, along the light blue line in Fig. \ref{Fig:fig1}(d) as a function of magnetic field $B_x$ along the [1$\bar{1}$0] crystallographic direction. At this plunger gate value, the dot has odd occupancy. The AC excitation voltage is $V_{AC}$ = 35 $\mu$V, while $QPC_L$ = -0.155 V and $QPC_R$ = -0.181 V. 
    b) Evolution of the differential conductance, with fixed plunger gate voltage $V_P$ = -0.5385 V, along the orange line in Fig. \ref{Fig:fig1}(d) as a function of magnetic field $B_x$ along the [1$\bar{1}$0] crystallographic direction. The quantum dot is now at even occupancy. The conduction may occur through either the spin-up or spin-down state of orbital level $n_2$.
    c) In-plane \textit{g}-factor for orbital levels $n_1$, $n_2$, and $n_3$. Green and red markers correspond to values of an \textit{g}-factor extracted from measurements with the magnetic field in the [1$\bar{1}$0] and [110] crystallographic directions, respectively. Triangles ($\blacktriangle$) and squares ($\blacksquare$) are extracted from the evolution of the zero-bias Coulomb blockade peaks in a magnetic field. The diamonds ($\blacklozenge$) and stars ($\bigstar$) correspond respectively to the differential conductance as a function of a magnetic field along the light blue and orange line cuts shown in the Coulomb blockade diamonds plot in Fig. \ref{Fig:fig1}(d).
    d) Evolution of the singlet GS and triplet ES along the green line shown in Fig. \ref{Fig:fig1}(d) as a function of the in-plane magnetic field $B_x$ along the [1$\bar{1}$0] crystallographic direction with a fixed number of electrons in the dot. The arrows along the conductance peaks represent the spin of the electron tunneling into a given orbital energy level. }
\label{Fig:fig3}
\end{figure}

\section{\label{results} Zeeman coupling with confinement}

One important parameter to understand in the confined quantum dot geometry is the effective \textit{g}-factor. Given the reduced symmetry of the dot, the \textit{g}-factor may vary with direction and may differ from values extracted in the planar 2DEG geometry. We explored Zeeman coupling with an in-plane magnetic field oriented parallel to the [1$\bar{1}$0] and [110] crystallographic directions. We also attempted to measure the out-of-plane response, but as will be discussed shortly, the lack of a clearly defined linear response regime prohibited identification of $|g^*_{[001]}|$.

The first method used for the extraction of the \textit{g}-factor analyzed the evolution of the Coulomb peaks at zero source-drain bias as a function of magnetic field \cite{kouwenhoven2001few, nilsson2009giant, kulesh2020quantum}.  Fig. \ref{Fig:fig2}(a) shows the evolution of the Coulomb peaks as a function of the in-plane magnetic field, $B_x$, along the [1$\bar{1}$0] direction. At zero magnetic field, sweeping the plunger towards less negative voltage adds electrons to the QD; each orbital energy level may be occupied by a spin-up and a spin-down electron. The conductance peaks correspond to a resonance where an electron is able to tunnel in and out of the dot, and the energy difference between consecutive peaks, known as the addition energy ($\Delta\mu$), can be calculated through the lever arm of the plunger gate as $\Delta\mu (B=0) = $ $\alpha_P\times$$\Delta V_P$. With the addition of magnetic field, each state has an additional Zeeman contribution to its energy: $\pm \frac{1}{2} g^* \mu_B B$. By analyzing the evolution of $\Delta\mu$ as a function of the magnetic field, we may extract the \textit{g}-factor. 

At low magnetic fields, $\Delta\mu (B)$ changes linearly due to the Zeeman energy, $E_z = g_n^*\mu_{B}B$. When two consecutive resonances move apart from one another, they correspond to spin-up ($\uparrow$), and spin-down ($\downarrow$) states belonging to the same orbital energy level. Fig. \ref{Fig:fig2}(b) shows the measured $\Delta\mu (B)$=$\Delta\mu (0) + E_z$ for three sets of consecutive transmission peaks belonging to three distinct orbital energy levels shown in Fig. \ref{Fig:fig2}(a), which we label as $n_1$, $n_2$, and $n_3$. Using a linear fit at a low magnetic field, we extract the \textit{g}-factor in the [1$\bar{1}$0] crystallographic direction, $|g^*_{[1\bar{1}0]}| \sim$ 55-59, the value varying slightly for the different orbital levels. Similarly, from the data shown in Fig. \ref{Fig:fig2}(c), the \textit{g}-factor in the [110] crystallographic direction is $|g^*_{[110]}| \sim$ 52-59. Fig. \ref{Fig:fig2}(d) shows the evolution of the Coulomb blockade peaks with the magnetic field $B_z$ parallel to the [001] crystallographic direction perpendicular to the plane of the quantum dot at low magnetic field; we do not observe a linear in magnetic field regime, precluding extraction of $|g^*_{[001]}|$.

With increasing magnetic field, every conductance resonance eventually undergoes a change in slope. This observation can be understood as a spin-down ($\downarrow$) state associated with a higher-lying orbital energy level lowering in energy due to the Zeeman contribution and passing through a lower-lying orbital state with spin-up ($\uparrow$). When a higher-lying orbital state with reversed spin becomes lower in energy than the lower-lying orbital state with spin-up ($\uparrow$), it becomes the new low-lying available state for conduction and the spin configuration of the dot is changed. This phenomenon is sketched in the schematic in Fig. \ref{Fig:fig2}(a). For example, the conductance resonance associated with the $n_2$ spin-down state (green arrow) changes its slope around $B_x$ = 0.12 T due to the spin-up state of the orbital energy level $n_3$ (light blue arrow) decreasing in energy, leading to a change in lowest lying available state in the dot. Accordingly, this change in slope of the third transmission peak from the top of Fig. \ref{Fig:fig2}(a) can be interpreted as a transition from a singlet to a triplet state, as shown by the green arrows in Fig. \ref{Fig:fig2}(a).  

The second method used to estimate the \textit{g}-factor of the confined electrons is through differential conductance measurements tracking the evolution of the QD's ground state as a function of in-plane magnetic field \cite{fasth2007direct, nilsson2009giant}. The conductance peaks delineating the Coulomb diamonds in Fig. \ref{Fig:fig1}(d) and intersected by the light blue and orange lines, reflect transport through the QD's ground state with an odd or even occupancy of electrons. The peaks occur when the electrochemical potential of either the source (-$V_{SD}$) or drain (+$V_{SD}$) aligns to the lowest unoccupied energy state in the QD. When transport is measured through the quantum dot occupied by an odd number of electrons, we measure the tunneling into a spin-down state, as shown in the schematic of Fig. \ref{Fig:fig3}(a) by the red arrow, while the transport through the dot with an even number of electrons corresponds to the tunneling of an electron to either of the available spin-split states, spin-up ($\uparrow$) and spin-down ($\downarrow$), of the respective orbital energy level as shown in Fig. \ref{Fig:fig3}(b) by the red arrows.  Fig. \ref{Fig:fig3}(a-b) shows transport evolving with magnetic field $B_x$ aligned to the [1$\bar{1}$0] crystallographic direction. The differential conductance measurements are also performed as a function of magnetic field in the [110] crystallographic direction and show nearly identical features as in the [1$\bar{1}$0] crystallographic direction.

Fig. \ref{Fig:fig3}(a) shows the evolution of the differential conductance at fixed plunger gate voltage of $V_P$ = -0.5385 V as a function of the magnetic field $B_x$ aligned to the [1$\bar{1}$0] crystallographic direction. Because the dot is at odd occupancy, we measure transport through the spin-down state ($\downarrow$) of orbital energy level $n_1$. The lowest energy differential conductance peaks increase towards higher $|V_{SD}|$ as a function of magnetic field since we have one unoccupied spin state associated with the orbital level $n_1$. The peak evolves as $E_Z$ $= g^*\mu_{B}B/2$. Due to the equal coupling of the source and drain contacts to the quantum dot, and the symmetric biasing of the dot, the energy is given by eV$_{SD}/2$. The extraction of the \textit{g}-factor in the [1$\bar{1}$0] crystallographic and in the [110] crystallographic direction yielded, $|g^*_{[1\bar{1}0]}|\sim$ 49 and $|g^*_{[110]}|\sim$ 49. The values are represented by the green and red diamonds in Fig. \ref{Fig:fig3}(c). 

The orange line in Fig. \ref{Fig:fig1}(d) corresponds to a fixed plunger gate voltage of $V_P$ = -0.5275 V. Fig. \ref{Fig:fig3}(b) shows transport through the quantum dot evolving with magnetic field $B_x$ along [1$\bar{1}$0] crystallographic direction. The lowest energy peaks measured at zero magnetic field correspond to the two available spin-up ($\uparrow$) and spin-down ($\downarrow$) states of the orbital energy level $n_2$. Each state evolves as a function of $E_Z$ $= g^*\mu_{B}B/2$, and the separation between the two peaks equals 2$E_Z$. The extracted \textit{g}-factors from the measurements shown in Fig. \ref{Fig:fig3}(b) are $|g^*_{[1\bar{1}0]}|\sim$ 55 and similarly  $|g^*_{[110]}|\sim$ 58, which are the green and red stars shown in Fig. \ref{Fig:fig3}(c). Both extraction methods described here yielded comparable values and show little anisotropy between the [1$\bar{1}$0] and [110] crystallographic directions.

 \begin{figure*}
    \includegraphics[width=1.0\textwidth]{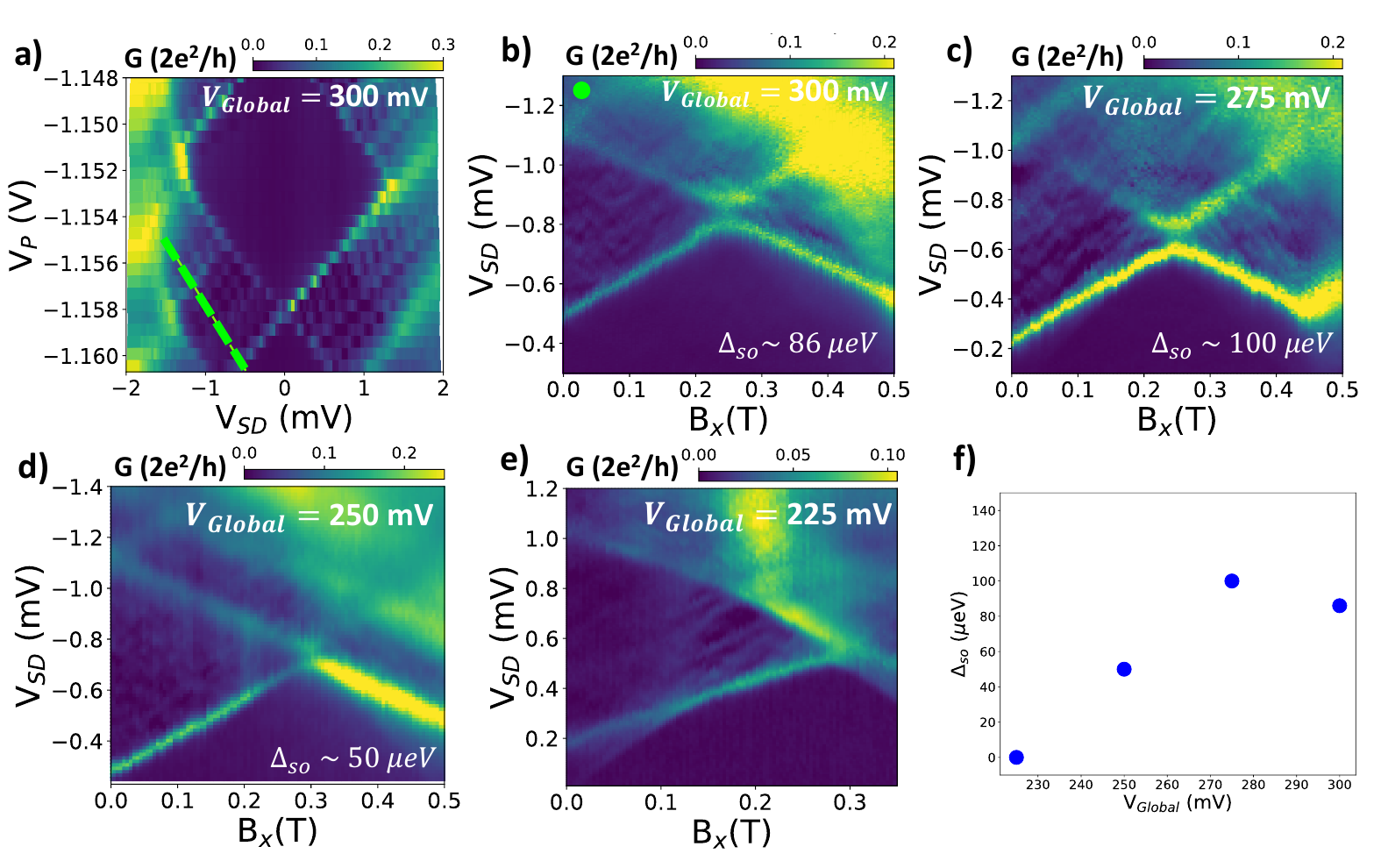}
    \caption{
        a) Charge stability diagram with +300 mV on the top global gate and tunneling barriers $QPC_{L}$ = $QPC_{R}$ = -0.255 V. $V_{AC}$ = 10 $\mu$V.
        b through e) Evolution of GS and ES with different gate voltages as a function of the in-plane magnetic field $B_x$ aligned to [1$\bar{1}$0] crystallographic direction. 
        b) Differential conductance measurement along the green line shown in Fig. 4(a). 
            $V_{Global}$ = + 300 mV, $V_{AC}$ = 6 $\mu$V, and  $QPC_{L}$ = $QPC_{R}$ = -0.255 V. 
        c) $V_{Global}$ = + 275 mV, $V_{AC}$ = 6 $\mu$V, and  $QPC_{L}$ = $QPC_{R}$ = -0.249 V. 
        d) $V_{Global}$ = + 250 mV, $V_{AC}$ = 6 $\mu$V, and  $QPC_{L}$ = -0.242V and $QPC_{R}$ = -0.245 V e) $V_{Global}$ = + 225 mV, $V_{AC}$ = 6 $\mu$V, and  $QPC_{L}$ = $QPC_{R}$ = -0.241 V.
        e) Spin-orbit mediated avoided crossing gap as a function of $V_{Global}$. 
}
    
    \label{Fig:fig4}
\end{figure*}

The \textit{g}-factor values extracted from the Coulomb blockade peaks at zero source-drain bias agree well with the values extracted from the differential conductance measurements. Similarly, both methods yield nearly isotropic \textit{g}-factors in the two crystallographic directions [1$\bar{1}$0] and [110]. Confined electrons in InSb$_{0.87}$As$_{0.13}$ quantum wells show a significantly higher in-plane \textit{g}-factor than that measured in quantum dots formed in a pure InSb quantum well of similar design, where $|g^*_{\parallel}|\sim$ 26-35 \cite{kulesh2020quantum}. Introduction of arsenic into the InSb lattice reduces the fundamental bandgap and results in enhanced interband matrix elements that increase the absolute value of the \textit{g}-factor. Overall, the value found in our experiment is slightly lower than the bulk \textit{g}-factor expected for InSb$_{0.87}$As$_{0.13}$, $|g| \sim 70$, estimated theoretically using the Roth formula \cite{roth1959theory, mayer2020superconducting}.  

\section{\label{SOC} Spin-Orbit-Coupling mediated avoided crossing}
Next, we study the spin-orbit coupling strength and measure the evolution of the ground state (GS) and an excited state (ES) as a function of the magnetic field. When SOC is treated perturbatively, the GS and an ES are mixed. In an applied magnetic field where spin selection rules are violated by SOC, this mixing yields level dispersion and an avoided crossing resolved as a gap between the energies of the GS and ES at finite in-plane field. The apparent gap $\Delta_{SO}$ is indicative of the strength of the mixing \cite{golovach2008spin}. 

Fig. \ref{Fig:fig3}(d) shows the differential conductance measurement along the green line in Fig. \ref{Fig:fig1}(d) as a function of magnetic field.  With this variation of $V_P$ and  $V_{SD}$, we probe the QD's GS and ES at an odd occupancy of electrons while the magnetic field is varied. In Fig. \ref{Fig:fig3}(d), we identify the lowest energy resonance as the singlet GS, with total spin $S$ = 0. This corresponds to the transport through the spin-down state of the orbital energy level $n_1$, as shown by the red arrow along the resonance. The higher energy resonances show transport through the ESs. The ES of two unpaired electrons, with total spin quantum number $S$ = 1,  splits into three triplet states, $T_+$, $T_0$, and $T_-$  with the spin's z-projection,  $S_z$ = + 1, 0, and -1 \cite{hanson2007spins}. While the GS increases in energy with magnetic field, the ES undergoes a splitting into two resonances. These resonances correspond to the transport of a spin-up and spin-down electron through the energy level $T_+$ ($S_z$ = 1) and $T_0$ ($S_z$ = 0). These transitions, $\uparrow \leftrightarrow T_+$ and $\uparrow \leftrightarrow T_0$, are understood as the two most favorable transitions \cite{fasth2007direct}. Fig. \ref{Fig:fig3}(d) shows the ES state, $T_+$, moving down in energy and closer to the GS, while the $T_0$ moves further away. $T_0$, indicated with a white dash line in Fig. \ref{Fig:fig3}(d), is resolved but has lower conductance due to the enhanced transition rate to the $T_+$ state. The measurements in Fig. \ref{Fig:fig3}(d) ($V_{Global}$ = + 50 mV) do not resolve a clear anti-crossing since the singlet GS intersects the $T_+$ ES,  indicating weak SOC at this particular value of chemical potential. At electron density $n \sim 1.6 \times 10^{11}$ cm$^{-2}$, self-consistent simulations \cite{Nextnano} of this heterostructure indicate a nearly symmetric quantum well and therefore weak SOC \cite{Metti2022}.

We increase the electric field to enhance the asymmetry of the quantum well and, consequentially, the Rashba coupling. Increasing $V_{Global}$ to +300 mV results in a density of $n = 2.7 \times 10^{11}$ cm$^{-2}$. As seen in Fig. \ref{Fig:fig4}(a), by remeasuring the evolution of a GS and ES as a function of the magnetic field at higher $V_{Global}$ along the green line in Fig. \ref{Fig:fig4}(a), we indeed observe a spin-orbit mediated avoided crossing between the singlet GS and the  ES, $T_+$, at around $B \sim$ 0.22 T. The avoided crossing gap extracted from spectroscopy is approximately $\Delta_{SO} \sim 86$ $\mu eV$. Under these circumstances, the SOC is enhanced due to the higher electric field across the heterostructure. The observed avoided crossing in a QD defined in an InSbAs 2DEG indicates gate control of the SOC strength. We measure the same GS and ES evolution as a function of the magnetic field at different $V_{Global}$ ranging from 225 mV to 300 mV, as seen in Fig. \ref{Fig:fig4} (b-e). As $V_{Global}$ decreases, the gap becomes smaller, indicating a decrease in SOC in agreement with Ref. \cite{Metti2022}. This is the first demonstration of a tunable SOC in a gate-defined quantum dot in this material system. The ability to tune spin-orbit coupling can provide utility for spin manipulation in mesoscopic devices. The maximum gap extracted from the avoided crossing is smaller than the one extracted in InSb nanowires of $\Delta_{SO} \sim 280$ $\mu eV$ \cite{nilsson2009giant,fan2015formation, mu2020measurements}; however, the measured gap depends significantly on the size of the dot, chemical potential, and the magnetic field \cite{golovach2008spin}, making direct comparison difficult.

\section{\label{conc}Conclusions}
In conclusion, we demonstrate the first realization of a tunable gate-defined quantum dot in a buried InSbAs quantum well with peak mobility of $\sim$ 200,000 cm$^2$/Vs. Our measurements indicate: (1) large in-plane \textit{g}-factor $\geq$ 50, and (2) {\it in-situ} tuning of SOC strength. These results should stimulate further investigation of this material system for spin-based devices and topological superconductivity. 

\nocite{*}

\bigskip
{$\dagger$} Author to whom correspondence should be addressed: mmanfra@purdue.edu

\bigskip
We thank Dr. James Nakamura for careful reading of this manuscript and valuable suggestions.

\bigskip
This work was supported by Microsoft Quantum. 

\bigskip

\bibliography{main}% Produces the bibliography via BibTeX.

\end{document}